\documentclass[preprint,aps,showpacs,prb,floatfix]{revtex4}
 
\usepackage{graphicx}
\usepackage{dcolumn}
\usepackage{bm}
 
\begin{document}

\preprint{Submitted to Phys. Rev. B}

\title{Electronic properties of the dimerized one-dimensional 
Hubbard model using lattice density-functional theory
      }
\author{R.~L\'opez-Sandoval}
\affiliation{Instituto Potosino de Investigaci\'on Cient\'{\i}fica 
y Tecnol\'ogica, Av.\ Venustiano Carranza 2425-A, 
78210 San Luis Potos\'{\i}, M\'exico}
\author{G.~M.~Pastor}
\affiliation{Laboratoire de Physique Quantique, Universit\'e Paul Sabatier, 
Centre National de la Recherche Scientifique, 118 route de Narbonne,
31062 Toulouse, France}

\date{\today}

\begin{abstract}

The dimerized one-dimensional Hubbard model is studied in the framework 
of lattice density-functional theory (LDFT). The single-particle density 
matrix $\gamma_{ij}$ with respect to the lattice sites is considered as
basic variable. The corresponding interaction-energy functional 
$W[\gamma_{ij}]$ is defined by Levy's constrained search. Exact 
numerical results are obtained for $W(\gamma_{12},\gamma_{23})$ where 
$\gamma_{12} = \gamma_{i,i+1}$ for $i$ odd and 
$\gamma_{23} = \gamma_{i,i+1}$ for $i$ even are the nearest-neighbor
density-matrix elements along the chain. The domain of representability 
of $\gamma_{ij}$ and the functional dependence of 
$W(\gamma_{12},\gamma_{23})$ are analyzed. A simple, explicit 
approximation to $W(\gamma_{12},\gamma_{23})$ is proposed, which is 
derived from scaling properties of $W$, exact dimer results, and known 
limits. Using this approximation, LDFT is applied to determine 
ground-state properties and charge-excitation gaps of finite and 
infinite dimerized chains as a function of the Coulomb-repulsion 
strength $U/t$ and of the alternation $\delta t$ of the hopping 
integrals $t_{ij}$ ($t_{ij} = t \pm \delta t$). The accuracy of the
method is demonstrated by comparison with available exact solutions and 
accurate numerical calculations. Goals and limitations of the present 
approach are discussed particularly concerning its ability to
describe the crossover from weak to strong electron correlations.

\end{abstract}
%


%
%

%
\maketitle

\section{\label{sec:introd}Introduction}

Hohenberg and Kohn (HK) replaced the wave function by the electronic
density $\rho(\vec r)$ as the fundamental variable of the many-body
problem and thereby achieved a crucial breakthrough in the theoretical
description of the electronic properties of matter.\cite{hk}
Since then, density functional theory (DFT) has been the subject
of remarkable developments. Formal improvements, extensions, 
and uncountable successful applications to a large variety 
of problems have made of this theory the most efficient, albeit 
not infallible, method of determining physical and chemical 
properties of matter from first principles.\cite{parr-book,gross-book}
Density functional (DF) calculations are usually based on the 
Kohn-Sham (KS) scheme which reduces the correlated $N$-particle 
problem to the solution of a set of self-consistent single-particle 
equations.\cite{ks} While this transformation is formally exact, the 
form of the interaction-energy functional $W[\rho(\vec{r})]$ involved 
in the KS equations is not known explicitly. Practical implementations 
of DFT always require approximations to $W[\rho(\vec{r})]$, or 
equivalently to the exchange and correlation (XC) functional 
$E_{\rm XC}[\rho(\vec{r})]$, on which the quality of the results depends 
crucially. Therefore, understanding the functional dependence of 
$W[\rho(\vec{r})]$ and exploring new ways of improving its 
approximations are central to the development of DF methods.

The most extensively used forms for $W[\rho(\vec{r})]$ 
are presently the local density approximation (LDA), \cite{ks}
its spin-polarized version or local spin-density approximation 
(LSDA)\cite{hedin} and the gradient corrected extensions,\cite{gc}
which were originally derived from exact results for the homogeneous 
electron gas. Despite an unparalleled success in the most diverse 
areas, the LDA-based approach fails systematically in accounting for
phenomena that involve strong electron-correlation effects 
as observed, for example, in Mott insulators, heavy-fermion 
materials, or high-$T_c$ superconductors. These systems are 
usually described in the framework of parametrized lattice 
Hamiltonians such as Anderson,\cite{andmod} Hubbard,\cite{hub} 
Pariser-Parr-Pople,\cite{ppp} and related models which focus 
on the most relevant electron dynamics at low energies. However,
even with simplified model interactions and a minimal number of 
orbitals per atom, a detailed understanding of the electronic properties
in the strongly correlated limit remains a serious theoretical challenge.
Exact results are rare or numerically very demanding and a variety 
elaborate many-body techniques have been specifically
developed in order to study this problem.\cite{fulde} 
Being in principle an exact theory, the limitations of the DF
approach have to be ascribed to the approximations used for 
exchange and correlation and not to the underlying formalism.
It is therefore very interesting to extend the range of
applicability of DFT to the many-body lattice models that describe 
the physics of strongly correlated systems. Moreover, the development 
of lattice density-functional theory (LDFT) constitutes an 
intrinsically inhomogeneous approach and provides a true 
alternative to the LSDA and related gradient-corrected methods.
Thus, studies on simple models can open new insights into the
properties of $W$ that should also be useful for future extensions 
to more realistic Hamiltonians and first principles calculations.

Several physical problems have been already investigated 
by applying the concepts of DFT to lattice models, for example, 
the band-gap problem in semiconductors,\cite{gunn} 
the role of off-diagonal elements of the density matrix 
and the non-interacting $v$ representability in strongly 
correlated systems,\cite{godby} or the development of energy 
functionals of the density matrix with applications to Hubbard 
and Anderson models.\cite{carl} 
In previous works\cite{prbxcfun,ldftdim} we have considered a 
density-matrix functional theory of many-body lattice models, 
that is analogous to Gilbert's approach in the continuum,\cite{gilb}
and applied it to the Hubbard Hamiltonian with uniform nearest-neighbor (NN)
hopping integrals $t_{ij} = t$. The interaction energy $W$ of the Hubbard 
model has been calculated exactly as a function of the density 
matrix $\gamma_{ij}$ for various periodic lattices having 
$\gamma_{ij} =\gamma_{12}$ for all nearest neighbors $i$ and $j$. 
An analysis of the functional dependence of $W(\gamma_{12})$ 
for different band fillings and lattice structures 
revealed very interesting scaling properties.\cite{prbxcfun} 
On this basis, a simple general approximation to $W(\gamma_{12})$ 
has been derived which yields a remarkable agreement with available 
exact results in 1D systems and which predicts successfully the 
ground-state energy and charge-excitation gap of the 2D Hubbard model 
in the complete range of interaction strength.\cite{ldftdim} 
This shows that DFT with an appropriate approximation to $W$ is an 
efficient tool for determining the electronic properties of many-body 
lattice models. 

The purpose of this paper is to extend the method by allowing for 
alternations of the density-matrix elements $\gamma_{ij}$ between 
nearest neighbors in order to study the dimerized 1D Hubbard model.
This problem has motivated a considerable research activity 
in past years, particularly concerning the role of electron correlations
in the dimerization of polymer chains like polyacetylene.\cite{teo-poly} 
In this context two qualitatively different regimes may be distinguished 
depending on the relative importance of the intra-atomic Coulomb 
repulsion $U$ and the NN hopping integral $t$.
On the one side, for small $U/t$, the dimerization can be regarded as
a bond-order wave that opens a gap at the Fermi surface
of the one-dimensional (1D) single-particle band-structure
(Peierls distortion). On the other side, for large $U/t$,
local charge fluctuations are severely reduced and the
low-energy properties are dominated by antiferromagnetic (AF)
correlations between spin degrees of freedom.
In this case the dimerization can be regarded as an
alternation of the strength of AF correlations along the chain
or spin-Peierls state. One of our aims is to analyze the differences
between these two types of behaviors in the framework of LDFT.
The properties of dimerized chains are also
very interesting from a purely methodological point of view.
They provide in fact a simple, physically motivated means of exploring
the functional dependence of $W[\gamma_{ij}]$ by including
additional degrees of freedom, thereby allowing for a larger flexibility.
Moreover, several exact results are available to compare with
(e.g., Bethe-Ansatz solution for the non-dimerized Hubbard chain,
finite-ring Lanczos diagonalizations, or density-matrix 
renormalization-group calculations) which allow to quantify 
the accuracy of the final results.

The remainder of the paper is organized as follows. In 
Sec.~\ref{sec:ldft} the main steps in the formulation of LDFT are 
briefly recalled. The properties of the interaction-energy 
functional $W$ of the dimerized Hubbard model are discussed
in Sec.~\ref{sec:W}. The domain of representability of 
$\gamma_{ij}$ and the scaling behavior of $W$ are investigated. 
A simple explicit approximation to $W$ is derived,
which is appropriate for direct calculations. Sec.~\ref{sec:dimhub} 
is concerned with applications. The ground-state energy and  
the charge-excitation gap of finite and infinite dimerized chains 
are determined as a function of Coulomb-repulsion strength $U/t$ and 
hopping-integral alternation $\delta t$. The LDFT results are 
contrasted with accurate numerical solutions in order to discuss 
goals and limitations of the present approach.
Finally, Sec.~\ref{sec:concl} summarizes the main conclusions
and points out some future perspectives.

\section{\label{sec:ldft}Density functional theory on a lattice}

We consider the many-body Hamiltonian 
\begin{equation}
\label{eq:hamgen}
H = \sum_{ij\sigma}
t_{ij} \; \hat c_{i \sigma}^{\dagger} \hat c_{j \sigma} +
\frac{1}{2} \sum_{klmn \atop{\sigma \sigma '}} 
V_{klmn} \; \hat c_{k \sigma }^{\dagger} \hat c_{m   \sigma '}^{\dagger} 
            \hat c_{n \sigma '} \hat c_{l \sigma} \;,
\end{equation}
where  $\hat c_{i \sigma}^{\dagger}$  ($\hat c_{i \sigma}$) is the 
usual creation (annihilation) operator for an electron with spin 
$\sigma$ at site or orbital $i$. The hopping integrals $t_{ij}$ 
define the lattice (e.g., 1D chains, square or triangular 2D 
lattices) and the range of single-particle interactions 
(e.g., up to first or second neighbors). From the {\em ab initio} 
perspective $t_{ij}$ is given by the external potential 
$V_{\rm ext}(\vec r)$ and by the choice of the basis. $V_{klnm}$ 
defines the type of many-body interactions which may be repulsive 
(Coulomb like) or attractive (in order to simulate electronic pairing) 
and which are usually approximated as short ranged (e.g.,
intra-atomic). Eq.~(\ref{eq:hamgen}) is mainly used in this section 
to present the general formulation which can then be applied to 
various specific models by simplifying the interactions. 
A particularly relevant example, to be considered in some detail in 
following sections, is the single-band Hubbard Hamiltonian with 
NN hoppings.\cite{hub} The non-dimerized form
of this model is obtained from Eq.~(\ref{eq:hamgen}) by setting 
$t_{ij}= -t$ for $i$ and $j$ NN's, $t_{ij} = 0$ otherwise, and 
$V_{klnm} = U \delta_{kl} \delta_{nm} \delta_{kn}$. 

The hopping matrix $t_{ij}$ plays the role given in conventional 
DFT to the external potential $V_{ext}(\vec r)$. Consequently, 
the single-particle density matrix $\gamma_{ij}$ between lattice 
sites replaces the continuum density $\rho(\vec r)$ as basic variable.
The situation is similar to the density-matrix functional theory 
proposed by Gilbert for the study 
of non-local pseudo-potentials 
$V_{\rm ext}(\vec r, \vec r')$.\cite{gilb,donnelly,val}
The ground-state energy $E_{gs}$ and density matrix $\gamma_{ij}^{gs}$
are determined by minimizing the energy functional
\begin{equation}
\label{eq:efun}
E[\gamma_{ij}] = E_K[\gamma_{ij}] + W [\gamma_{ij}]
\end{equation}
with respect to $\gamma_{ij}$. $E[\gamma_{ij}]$ is defined for all 
density matrices that can be written as
\begin{equation}
\label{eq:gamij} 
\gamma_{ij} = \sum_{\sigma}\gamma_{ij \sigma}=
\sum_\sigma \langle \Psi| \hat c_{i \sigma }^{\dagger}
                          \hat c_{j \sigma } |\Psi \rangle 
\end{equation}
for all $i$ and $j$, where $|\Psi \rangle$ is an $N$-particle state. 
In other words, $\gamma_{ij}$ derives from a physical state and is 
said to be pure-state $N$-representable.\cite{foot_ens} The first 
term in Eq.~(\ref{eq:efun}) is given by
\begin{equation}
\label{eq:ek}
E_K = \sum_{ij} t_{ij}\gamma_{ij} \; .
\end{equation}
It represents the kinetic energy associated with the electronic 
motion in the lattice and includes all single-particle contributions. 
Notice that Eq.~(\ref{eq:ek}) yields the exact kinetic energy and
that no corrections on $E_K$ have to be included in other parts of 
the energy functional as in the KS approach. The second term in 
Eq.~(\ref{eq:efun}) is the interaction-energy 
functional given by\cite{levy}
\begin{equation}
\label{eq:w}
W[ \gamma_{ij}] = 
min \left[\frac{1}{2} \sum_{ klmn \atop {\sigma \sigma '} }  
V_{klmn} \; \langle \Psi [ \gamma_{ij} ] | \;
\hat c_{k \sigma }^{\dagger} \hat c_{m \sigma '}^{\dagger}
\hat c_{n \sigma '}          \hat c_{l \sigma} 
\; |\Psi[\gamma_{ij}]\rangle  \right] \; ,
\end{equation}
where the minimization implies a search over all 
$N$-particles states $| \Psi [\gamma_{ij}] \rangle$ 
that satisfy 
\begin{equation}
\label{eq:rep}
\langle \Psi [\gamma_{ij}] | \; 
\sum_\sigma \hat c_{i \sigma }^{\dagger}\hat c_{j \sigma} \; 
|\Psi [\gamma_{ij}] \rangle = \gamma_{ij} 
\end{equation}
for all $i$ and $j$. $W[\gamma_{ij}]$ represents the 
minimum value of the interaction energy compatible with a given 
density matrix $\gamma_{ij}$. It is often expressed in terms of 
the Hartree-Fock energy 
\begin{equation}
\label{eq:HF}
E_{\rm HF}[\gamma_{ij}]=\frac{1}{2}\sum_{ijkl \atop{\sigma \sigma'}} 
V_{ijkl} \left(\gamma_{ij \sigma} \gamma_{kl \sigma'} 
- \delta_{\sigma\sigma '}  \gamma_{il \sigma} \gamma_{kj \sigma} 
\right) 
\end{equation}
\noindent and the correlation energy $E_{\rm C}[\gamma_{ij}]$ as
\begin{equation}
\label{eq:xc}
W[\gamma_{ij}] = E_{\rm HF}[\gamma_{ij}] + E_{\rm C}[\gamma_{ij}] \;.
\end{equation}                  
$W$ and $E_{\rm C}$ are universal functionals of $\gamma_{ij}$ in 
the sense that they are independent of $t_{ij}$, i.e., of the 
system under study. They depend on the considered interactions or 
model, as defined by $V_{klmn}$, on the number of electrons $N_e$,
and on the structure of the many-body Hilbert space, as given by 
$N_e$ and the number of orbitals or sites $N_a$. 

$E[\gamma]$ is minimized by expressing 
$\gamma_{ij} = \gamma_{ij\uparrow} + \gamma_{ij\downarrow}$ 
in terms of the eigenvalues $\eta_{k\sigma}$ (occupation 
numbers) and eigenvectors $u_{ik\sigma}$ (natural orbitals) as 
\begin{equation}
\label{eq:gamma}
\gamma_{ij\sigma} =  \sum_k u_{ik\sigma} \;  \eta_{k\sigma} \;
u_{jk\sigma}^* \; .
\end{equation}
Lagrange multipliers $\mu$ and $\lambda_{k\sigma}$ 
($\varepsilon_{k\sigma} = \lambda_{k\sigma} / \eta_{k\sigma}$)
are introduced in order to impose the usual constraints
$\sum_{k\sigma} \eta_{k\sigma} = N_e$ and
$\sum_i\vert u_{ik\sigma}\vert^2 = 1$.
Derivation with respect to $u_{jk\sigma}^*$ and $\eta_{k\sigma}$
($0 \le \eta_{k\sigma}\le 1$) yields the eigenvalue 
equations\cite{ldftdim,gilb}
\begin{equation}
\label{eq:minsc}
\eta_{k\sigma}
\sum_i \left(t_{ij} + {\partial W \over \partial \gamma_{ij}} \right)
u_{ik\sigma} = \varepsilon_{k\sigma} u_{jk\sigma}
\end{equation}
with the following conditions relating $\eta_{k\sigma}$ and
$\varepsilon_{k\sigma}$:
\begin{eqnarray}
\label{eq:minepsa}
\varepsilon_{k\sigma} & < & \mu \quad  {\rm if} \quad \eta_{k\sigma} = 1\; ,\\
\label{eq:minepsb}
\varepsilon_{k\sigma} & = & \mu \quad {\rm if} \quad
0<\eta_{k\sigma}< 1\; ,
\end{eqnarray}
and
\begin{eqnarray}
\label{eq:minepsc}
\varepsilon_{k\sigma} & > & \mu \quad  {\rm if} \quad \eta_{k\sigma} = 0\; .
\end{eqnarray}
Self-consistency is implied by the dependence of
$\partial W / \partial\gamma_{ij}$ on $\eta_{k\sigma}$ and
$u_{ik\sigma}$. Eqs.~(\ref{eq:minsc})--(\ref{eq:minepsc})
hold exactly in all interaction regimes. They are analogous to 
well-known results of density-matrix functional theory in the
continuum.\cite{gilb} However, notice the difference with
the KS-like approach considered in Ref.~\cite{godby}, which 
assumes non-interacting $v$-representability, and where 
only integer occupations are allowed. 

The importance of fractional natural-orbital occupations 
has already been stressed in previous density-matrix functional studies
in the continuum.\cite{gilb} In fact, in the case of models 
one observes that $0<\eta_{k\sigma}<1$ for all $k\sigma$ 
except in very special situations like the uncorrelated
limit ($V_{klmn} = 0$) or the fully-polarized ferromagnetic state in the 
Hubbard model. This can be understood from perturbation-theory 
arguments ---none of the $\eta_{k\sigma}$ should be a
good quantum number for $V_{klmn}\not= 0$--- and has been explicitly
demonstrated in exact solutions for finite systems or for the
1D Hubbard chain.\cite{lieb-wu} Therefore, the case (\ref{eq:minepsb})
is the only relevant one in general. All $\varepsilon_{k\sigma}$
in Eq.~(\ref{eq:minsc}) must be degenerate or equivalently
\begin{equation}
\label{eq:tij}
t_{ij} + \partial W / \partial\gamma_{ij \sigma} = \delta_{ij} \; \mu \;.
\end{equation}
Clearly, approximations of $W$ in terms of the diagonal $\gamma_{ii}$ 
alone can never yield such a behavior. Given a self-consistent 
scheme that implements the variational principle, the challenge remains 
to find good approximations to $W[\gamma_{ij}]$ that are simple enough 
to be applied in practical calculations.

\section{\label{sec:W}Interaction-energy functional}

In order to determine $W[\gamma_{ij}]$ from Eq.~(\ref{eq:w}) 
we look for the extremes of
\begin{eqnarray}
\label{eq:f}
F &=& \sum_{ klmn \atop {\sigma \sigma '} }  \left [ V_{klmn}  
\langle\Psi| \hat c_{k \sigma }^{\dagger} \hat c_{m \sigma '}^{\dagger}
\hat c_{n \sigma } \hat c_{l \sigma} |\Psi\rangle \right]
 \; + \;\varepsilon \; \Big(1 - \langle \Psi |\Psi \rangle \Big) 
 \;  \nonumber  \\
 &+& \sum_{i,j} \lambda_{ij} \;
 \Big( \langle\Psi| 
       \sum_\sigma \hat c^\dagger_{i\sigma} \hat c_{j \sigma} |
       \Psi\rangle  -  \gamma_{ij} \Big) 
\end{eqnarray}
with respect to $|\Psi\rangle $. Lagrange multipliers $\varepsilon$ 
and $\lambda_{ij}$ have been introduced to enforce the normalization of 
$|\Psi\rangle$ and the representability of $\gamma_{ij}$ as required
by Eq.~(\ref{eq:rep}). Derivation with respect to $ \langle\Psi|$, 
$\varepsilon$ and $\lambda_{ij}$ yields the eigenvalue 
equations\cite{prbxcfun} 
\begin{equation}
\label{eq:evgen}
\sum_{ij\sigma} \lambda_{ij} \; \hat c^{\dagger}_{i \sigma} \hat c_{j \sigma} 
\; |\Psi \rangle  + 
\sum_{ klmn \atop {\sigma \sigma '} } V_{klmn}\;  
\hat c_{k \sigma }^{\dagger} \hat c_{m \sigma '}^{\dagger}
\hat c_{n \sigma } \hat c_{l \sigma} \; |\Psi\rangle  
= \varepsilon \; | \Psi \rangle \; , 
\end{equation}
and the auxiliary conditions $\langle \Psi | \Psi \rangle = 1$ and 
 $\gamma_{ij} = \langle\Psi| 
\sum_\sigma \hat c^\dagger_{i\sigma} \hat c_{j \sigma} | \Psi\rangle$. 
The Lagrange multipliers $\lambda_{ij}$ play the role of hopping
integrals to be chosen in order that $|\Psi\rangle$ yields the given 
$\gamma_{ij}$. The pure-state representability of $\gamma_{ij}$
ensures that there is always a solution. The subset of $\gamma_{ij}$ 
that can be represented by a ground-state of Eq.~(\ref{eq:evgen}) 
for some $\lambda_{ij}$ is the physically relevant one, since it 
necessarily includes the absolute minimum $\gamma_{ij}^{gs}$ of 
$E[\gamma_{ij}]$. Nevertheless, it should be noted that pure-state 
representable $\gamma_{ij}$ may be considered that can only be described 
by excited states or by linear combinations of eigenstates 
of Eq.~(\ref{eq:evgen}).\cite{prbxcfun}

The general functional $W[\gamma_{ij}]$, valid for all lattice structures 
and for all types of hybridizations, can be simplified at the expense of 
universality if the hopping integrals are short ranged. For example, 
if only NN hoppings are considered, the kinetic energy $E_K$ is 
independent of the density-matrix elements between sites that are not NN's. 
Therefore, the constrained search in Eq.~(\ref{eq:w}) may be restricted 
to the $| \Psi [\gamma_{ij}] \rangle$ that satisfy Eq.~(\ref{eq:rep})
only for $i=j$ and for NN $ij$. This reduces significantly the number 
of variables in $W[\gamma_{ij}]$ and renders the determination and
interpretation of the functional dependence far simpler. 

In Sec.~\ref{sec:Wex} we present and discuss exact results for the interaction 
energy $W[\gamma_{ij}]$ of the dimerized Hubbard model on representative 
finite and infinite chains. These are obtained by solving 
Eq.~(\ref{eq:evgen}) using accurate numerical methods. The dependence 
of the interaction energy on the alternating NN density-matrix elements 
$\gamma_{12}$ and $\gamma_{23}$ is analyzed. Scaling properties are 
identified within the domain of representability of $\gamma_{ij}$. 
On the basis of these results we propose, in Sec.~\ref{sec:Wdim}, a 
simple general approximation to $W(\gamma_{12},\gamma_{23})$ which is
useful for practical applications. A first test on the accuracy of this 
approximation is also provided by comparison with available exact solutions.

\subsection{\label{sec:Wex}Exact calculated $W[\gamma_{ij}]$
of the dimerized Hubbard model} 

In the following we consider the dimerized 1D Hubbard model which 
in the usual notation is given by\cite{hub}
\begin{equation}
\label{eq:hamhub}
H = \sum_{\langle ij\rangle \sigma} t_{ij}
\hat c^{\dagger}_{i \sigma} \hat c_{j \sigma} +
U \sum_i  \hat n_{i \downarrow} \hat n_{i\uparrow} \;. 
\end{equation}
The NN hopping integrals $t_{ij}$ take two alternating values:
$t_{i,i+1} = t_{12} = t + \delta t$ for $i$ odd and
$t_{i,i+1} = t_{23} = t - \delta t$ for $i$ even.
The corresponding interaction-energy functional reads
\begin{equation}
\label{eq:whub}
W[\gamma_{ij}] = 
min \left[ U \sum_l  \langle \Psi [\gamma_{ij}] |\; 
\hat n_{l\uparrow} \hat n_{l\downarrow} 
\;|\Psi [\gamma_{ij}] \rangle  \right] \; ,
\end{equation}
where the minimization is performed with respect to all $N$-particle states 
$|\Psi [\gamma_{ij}]\rangle$ satisfying $\langle \Psi [\gamma_{ij}] | 
\sum_\sigma \hat c_{i \sigma }^{\dagger}\hat c_{j \sigma} |
\Psi [\gamma_{ij}] \rangle = \gamma_{ij}$ 
for NN $ij$. For repulsive interactions $W[\gamma_{ij}]$ represents 
the minimum average number of double occupations corresponding to a 
given degree of electron delocalization, i.e., to a given 
$\gamma_{ij}$. Eq.~(\ref{eq:evgen}) then reduces to 
\begin{equation}
\label{eq:evhub}
\sum_{\langle ij\rangle \atop \sigma} \lambda_{ij} \;
\hat c^{\dagger}_{i\sigma}\hat c_{j\sigma} \; |\Psi\rangle \; + \;
U \sum_i \hat n_{i\uparrow} \hat n_{i\downarrow} \; |\Psi\rangle  
= \varepsilon  \; |\Psi\rangle \; .
\end{equation}
This eigenvalue problem can be solved numerically for finite systems
with various boundary conditions. To this aim
we expand $| \Psi [\gamma_{ij}] \rangle$
in a complete set of basis states $\vert\Phi_m\rangle$ which have 
definite occupation numbers $\nu^m_{i \sigma}$ at all orbitals $i\sigma$: 
$\hat n_{i\sigma} \vert\Phi_m\rangle = \nu_{i\sigma}^m 
\vert\Phi_m\rangle$ 
with $\nu_{i\sigma}^m = 0$ or $1$. The values of $\nu^m_{i \sigma}$ 
satisfy the usual conservation of the number of electrons 
$N_e  = N_{e\uparrow} + N_{e\downarrow}$ and of the $z$ component 
of the total spin $S_z = (N_{e\uparrow} - N_{e\downarrow})/2$, where
$N_{e\sigma} = \sum_{i} \nu^m_{i \sigma}$. For not too large clusters,
the state $|\Psi_0 [\gamma_{ij}] \rangle$ corresponding to the minimum
in Eq.~(\ref{eq:whub}) ---the ground state of Eq.~(\ref{eq:evhub})--- 
can be determined by sparse-matrix diagonalization procedures such as 
the Lanczos iterative method.\cite{lanczos}
For large chains, the properties of $|\Psi_0 [\gamma_{ij}] \rangle$ 
can be calculated using the
density-matrix renormalization-group (DMRG) method\cite{white}
which allows reliable extrapolations to the infinite-length limit. 
Finally, in the absence of dimerization 
($\delta t = 0$) translational symmetry implies that all NN 
$\gamma_{ij}$ are the same, and therefore one may set 
$\lambda_{ij} = \lambda$ for all NN $ij$. The lowest eigenvalue 
of Eq.~(\ref{eq:evhub}) can then be determined 
from Lieb and Wu's exact solution of the 1D Hubbard model 
following the work by Shiba.\cite{lieb-wu} 

\begin{figure}
\includegraphics[scale=0.7]{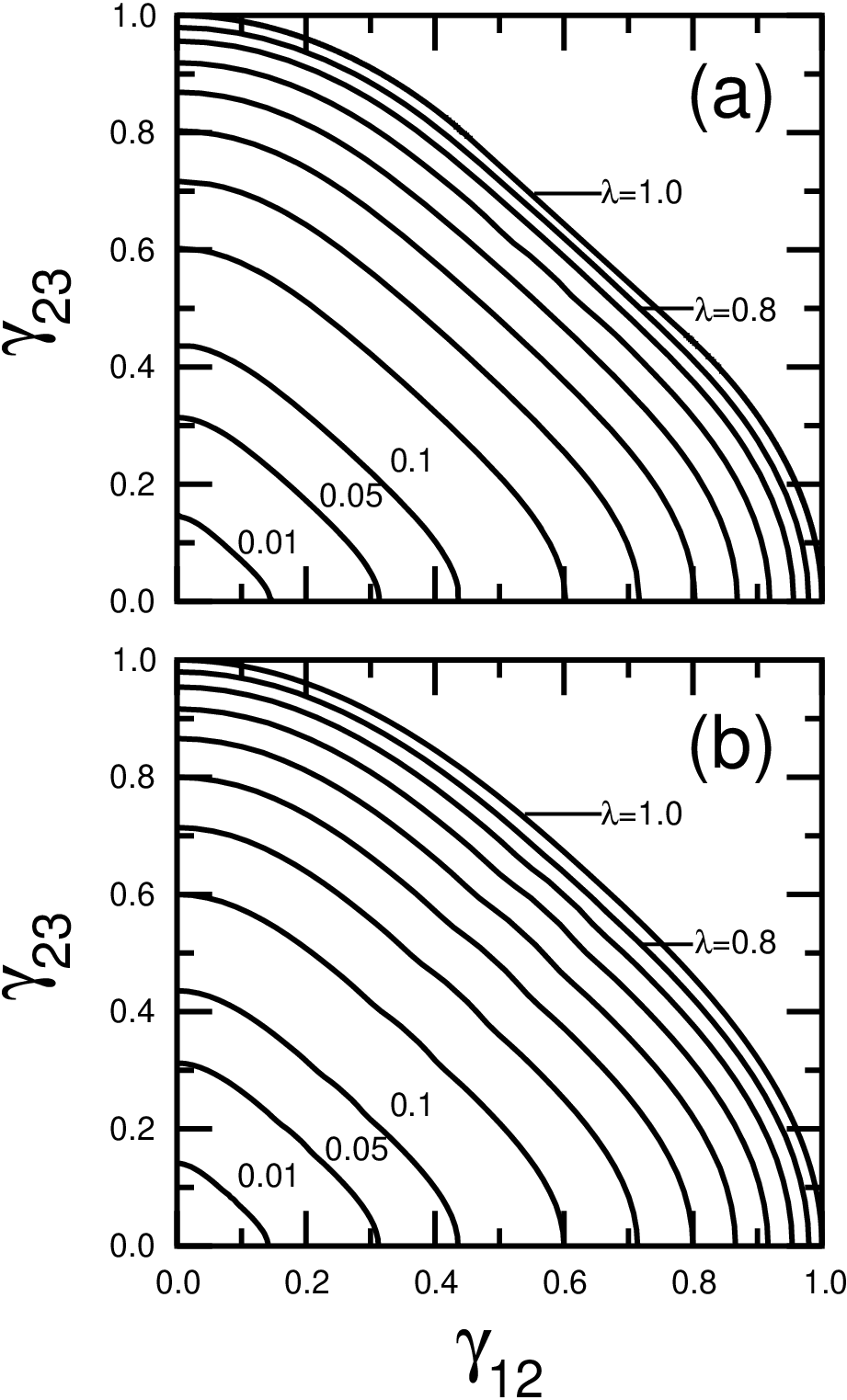}
\caption{\label{fig:Wniv}
Constant interaction-energy curves of the one-dimensional (1D)
Hubbard model as given by 
$W(\gamma_{12}, \gamma_{23}) = \lambda E_{\rm HF}$, where 
$E_{\rm HF} = U/4$ is the Hartree-Fock energy and $\lambda$ a constant
($0 \le \lambda \le 1$). 
The NN density-matrix elements are $\gamma_{i,i+1} = \gamma_{12}$ 
for $i$ odd, and $\gamma_{i,i+1} = \gamma_{23}$ for $i$ even. 
Results are given for (a) the $N_a = 12$ 
site ring and (b) the infinite chain, both at half-band filling 
($N_e = N_a$). $\lambda = 1$ corresponds to uncorrelated states
and defines the limit of representability of $\gamma_{ij}$.
Unless indicated the difference in $\lambda$ between contiguous 
curves is $\Delta\lambda = 0.1$.
        }
\end{figure}

In Fig.~\ref{fig:Wniv} the interaction energy $W$ of dimerized 
Hubbard chains is shown in the form of constant-energy curves 
given by $W(\gamma_{12}, \gamma_{23}) = \lambda E_{\rm HF}$,
where $\gamma_{12} = \gamma_{i,i+1}$ for $i$ odd and 
$\gamma_{23} = \gamma_{i,i+1}$ for $i$ even are the density-matrix 
elements or bond orders between NN's. $E_{\rm HF}=  U/4$ stands for the 
Hartree-Fock energy, and $\lambda$ is a constant ($0\le \lambda\le 1$). 
Results are presented for the $N_a = 12$ site ring and for the 
infinite 1D chain which were obtained from Eq.~(\ref{eq:evhub}) 
by using Lanczos-diagonalization and DMRG methods, 
respectively.\cite{lanczos,white} Only positive $\gamma_{12}$ and 
$\gamma_{23}$ are considered since this is the relevant domain
when all the hopping integrals have the same sign. In bipartite lattices, 
like open chains or rings with even $N_a$, the sign of the NN bond orders 
can be changed without altering $W$ by changing the sign of the 
local orbitals at one of the sublattices. Thus,  
$W(\gamma_{12}, \gamma_{23}) = W(-\gamma_{12}, -\gamma_{23})$.
Moreover, $W(\gamma_{12}, \gamma_{23}) = W(\gamma_{23}, \gamma_{12})$
as even an odd sites can be interchanged by a simple translation
($N_a$ even for rings). 

The domain of definition of $W$ is restricted by the pure-state 
representability of $\gamma_{ij}$. 
The axes $\gamma_{12} = 0$ and $\gamma_{23} = 0$ in Fig.~\ref{fig:Wniv}
represent a collection of disconnected dimers or fully dimerized states, 
while $\gamma_{12} = \gamma_{23}$ corresponds to non-dimerized 
states. In between, the degree of dimerization can be 
measured by the angle $\phi= \arctan(\gamma_{12}/\gamma_{23})$. 
The degree of electron delocalization for each $\phi$ is
characterized by $\gamma = \sqrt{\gamma_{12}^2 + \gamma_{23}^2}$,
which is bounded by $\gamma^\infty(\phi) \le \gamma \le \gamma^0(\phi)$
in order that $\gamma_{ij}$ remains pure-state representable.
The density-matrix elements along the curve $\gamma = \gamma^0(\phi)$
are the largest bond orders that can be achieved  
on a given lattice and for given $N_a$ and $N_e$
[$(\gamma_{12}^0, \gamma_{23}^0) = \gamma^0(\phi) (\cos\phi, \sin\phi)$]. 
They represent the maximum electron delocalization for 
each $\phi$ and yield the extremes of the kinetic energy 
$E_K=\sum t_{ij}\gamma_{ij}$, with different $\phi$ corresponding
to different $t_{12}/t_{23}$. Thus, for $\gamma = \gamma^0(\phi)$
the density matrix can be represented by the ground state
of the uncorrelated Hubbard model for some $t_{12}/t_{23}$ ($U=0$).
In the absence of degeneracy the underlying electronic state 
$|\Psi_0 \rangle$ is a single Slater determinant and 
$W(\gamma_{12}^0, \gamma_{23}^0) = E_{\rm HF}$. Consequently, 
the upper bound for
$\gamma$ coincides with the $\lambda = 1$ curve in Fig.~\ref{fig:Wniv}. 
The correlation energy $E_C=W-E_{\rm HF}$ vanishes as expected in 
the fully delocalized limit. For $U=0$ the minimization of the 
energy $E=E_K$ as a function of $\gamma_{ij}$ can be stated in terms of 
the representability of $\gamma_{ij}$ alone. In this case 
the equilibrium condition yielding $\gamma_{ij}^{gs}$ 
is achieved at the borders of the domain of representability, 
more precisely, when the normal to the curve $\gamma = \gamma^0(\phi)$ 
is parallel to $\vec\nabla E_K = (t_{12}, t_{23})$.

Concerning the lower bound $\gamma^\infty(\phi)$, one should first note 
that as $\gamma$ decreases, $\gamma < \gamma^0(\phi)$, it is possible 
to construct correlated states $|\Psi[\gamma_{ij}] \rangle$ having  
increasingly localized electrons. Charge fluctuations can then be 
reduced more efficiently for smaller $\gamma$, and therefore the Coulomb 
interaction energy decreases with decreasing $\gamma$
[see Eq.~(\ref{eq:whub}) and Fig.~\ref{fig:Wniv}]. $W$ reaches its 
minimum value $W_\infty = U max\{0,N_e-N_a\}$ in the strongly correlated 
limit where $\gamma = \gamma^\infty(\phi)$. For half-band 
filling this corresponds to a fully localized state
having $\gamma^\infty(\phi) = 0$ and $W_\infty=0$. 
However, note that for $N_e \not= N_a$, $W$ reaches $W_\infty$ already
for $\gamma^\infty(\phi) > 0$ since partially 
delocalized states can be found having minimal Coulomb repulsion. 
This is the case for example in a fully-polarized ferromagnetic state.

Fig.~\ref{fig:Wniv} also provides a qualitative picture of the functional 
dependence of $W$ for dimerized chains.  On the one side, for strongly 
dimerized states ($\phi\simeq 0$ or $\phi\simeq \pi/2$) the constant-$W$ 
curves resemble circumference arcs, the gradient $\vec\nabla W$ being 
approximately radial. This type of behavior is most clearly seen 
for weak or moderate correlations ($\lambda \ge 0.3$), while in 
the localized regime ($\lambda \le 0.1$) it holds only for a very 
limited range of $\phi$ around $\phi = 0$ or $\phi = \pi/2$.
On the other side, for weakly to moderately dimerized states 
($\phi\simeq \pi/8$--$\pi/4$) the level curves can be regarded 
in first approximation as straight lines parallel to 
$\gamma_{12} = - \gamma_{23}$. The very weak dependence of $\vec\nabla W$
on $\phi$ implies that for $\phi\simeq \pi/4$ the ground-state values 
of $\gamma_{12}^{gs}$ and $\gamma_{23}^{gs}$, which result from the 
minimization of $E = E_K + W$, are very sensitive to the hopping 
alternation $\delta t$. In fact, 
significant variations of $\phi$ are necessary until 
$\vec\nabla W = - \vec\nabla E_K \propto (1+\delta t /t, 1-\delta t /t)$
even for $\delta t /t \ll 1$. This is particularly notable for 
weak correlations since the $W=E_{\rm HF}$ curve is strictly linear
for $|\phi-\pi/4| < 0.05$. Therefore, a discontinuous change
from $\gamma_{12}^{gs} / \gamma_{23}^{gs} = 1$ to 
$\gamma_{12}^{gs} / \gamma_{23}^{gs} = 0.91$
is found at $U=0$ and arbitrary small $\delta t$. For $U>0$, 
$\gamma_{12}^{gs}$ and $\gamma_{23}^{gs}$ are continuous functions of 
$\delta t$, although the dependence on $\delta t$ remains 
very strong for small $\delta t$, as can be inferred from the 
level curves in the figure.
Comparing subfigures (a) and (b) one observes that the 
results for $N_a = 12$ and $N_a = \infty$ are quite similar. 
The rather rapid convergence with chain length suggests
that $W(\gamma_{12}, \gamma_{23})$ is not very sensitive 
to the details of the considered system, even if the minimization 
constraints in Eq.~(\ref{eq:whub}) apply only to NN bond orders. This 
is of interest for practical applications, as it will 
be discussed below.

\begin{figure}
\includegraphics[scale=0.7]{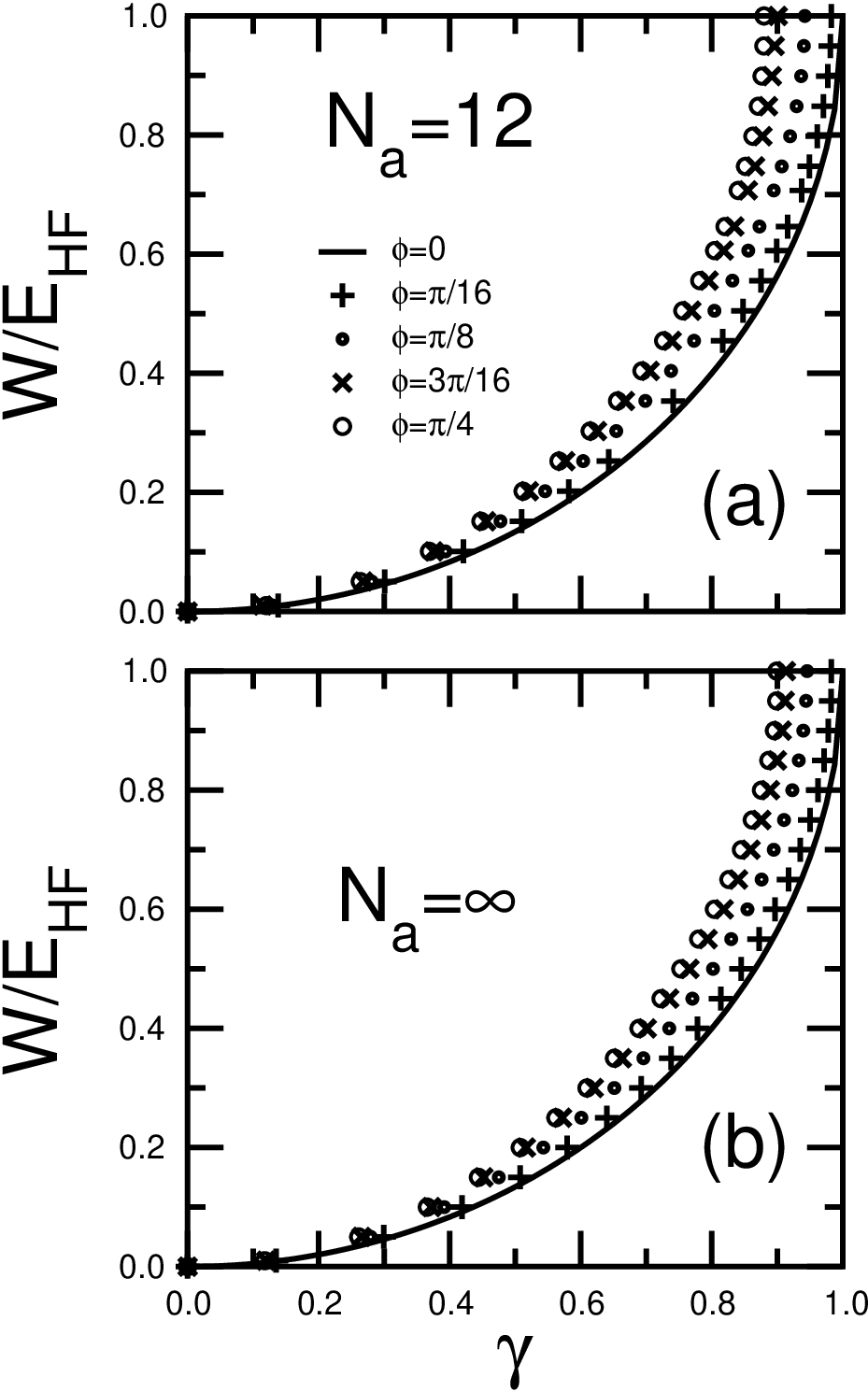}
\caption{\label{fig:Wgamma}
Interaction-energy $W$ of the 1D
Hubbard model at half-band filling ($N_e = N_a$) as a function of 
$\gamma = \sqrt{\gamma_{12}^2 + \gamma_{23}^2}$ for different values of
$\phi = \arctan(\gamma_{12}/\gamma_{23})$:
(a) ring with $N_a = 12$ sites (b) infinite chain.        
The density-matrix elements are
$\gamma_{i,i+1} = \gamma_{12}$ for $i$ odd, and 
$\gamma_{i,i+1} = \gamma_{23}$ for $i$ even.
        }
\end{figure}

In Fig.~\ref{fig:Wgamma} the interaction energy $W$ is shown as a 
function of $\gamma$ for representative values of $\phi$, including 
in particular the fully-dimerized ($\phi=0$) and non-dimerized
($\phi=\pi/4$) cases. Despite the quantitative differences
among the various $\phi$, several qualitative properties are 
shared by all the curves:
(i) As already discussed, the domain of representability
of $\gamma$ is bound for each $\phi$ by the bond order 
$\gamma^0(\phi)$ in the uncorrelated limit. 
$\gamma^0$ decreases monotonously with increasing
$\phi$ for $0 \le \phi \le \pi/4$ showing that a compromise 
between $\gamma_{12}$ and $\gamma_{23}$ is made when the 
two bonds are active. This is an important contribution 
to the $\phi$-dependence of $W$.
(ii) In the delocalized limit, $W(\gamma^0,\phi) = E_{\rm HF} = U/4$
for all $\phi$, since the electronic state yielding the largest 
$\gamma$ is a single Slater determinant. Moreover, one observes 
that $\partial W / \partial\gamma$ diverges at $\gamma = \gamma^0$ 
for all $\phi$. This is a necessary condition in order that 
the ground-state density matrix satisfies $\gamma^{gs} < \gamma^0$ 
for arbitrary small $U>0$, as expected from perturbation theory.
(iii) Starting from $\gamma = \gamma^0$, $W$ decreases with 
decreasing $\gamma$ reaching its lowest possible value $W=0$ for 
$\gamma = 0$ ($N_e = N_a$). The decrease of $W$ with decreasing 
$\gamma$ means that the reduction of the Coulomb energy due
to correlations is done at the expense of kinetic energy or
electron delocalization.
(iv) In the limit of small $\gamma$ one observes that  
$W \propto \gamma^2$. Therefore, for $U/t\gg 1$, 
$\gamma^{gs} \propto t/U$ and $E_{gs} \propto t^2/U$, 
a well known result in the Heisenberg limit of the half-filled Hubbard 
model.\cite{fulde}

\subsection{\label{sec:Wdim}Scaling approximation to $W[\gamma_{ij}]$} 

\begin{figure}
\includegraphics[scale=0.7]{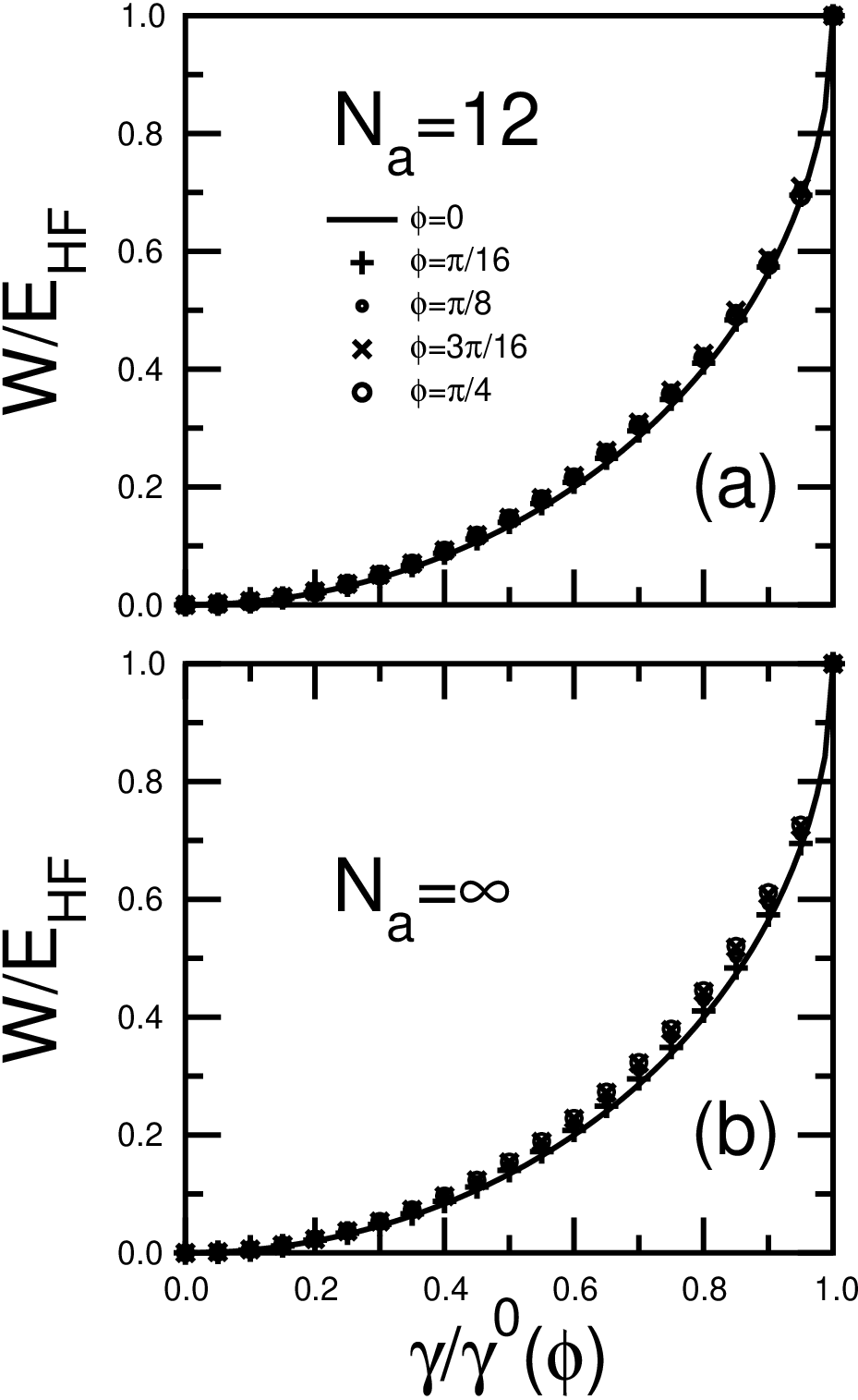}
\caption{\label{fig:Wgren}
Interaction-energy $W$ of the 1D Hubbard model as a function 
of $\gamma / \gamma^0$ 
for different $\phi = \arctan(\gamma_{12}/\gamma_{23})$.
$\gamma = \sqrt{\gamma_{12}^2 + \gamma_{23}^2}$ and $\gamma^0(\phi)$ 
is the largest representable value of $\gamma$ for the given $\phi$,
which corresponds to the uncorrelated limit ($0\le \gamma\le \gamma^0$, 
see Fig.~\protect\ref{fig:Wniv}). Results are shown for 
(a) the $N_a = 12$ site ring and (b) the infinite chain, both at 
half-band filling.        
        }
\end{figure}

In order to compare the $\gamma$-dependence of $W$ for different 
$\phi$ and to analyze its scaling behavior it is useful to bring the 
domains of representability for different $\phi$ to a common range 
by considering $W(\gamma,\phi)$ as a function of $\gamma / \gamma^0(\phi)$, 
as displayed in Fig.~\ref{fig:Wgren}. In this form the results for
different $\phi$ appear as remarkably similar, showing that the 
largest part of the dependence of $W$ on the ratio 
$\gamma_{12} / \gamma_{23}$ comes from the domain of representability 
of $\gamma_{ij}$ given by its upper bound $\gamma^0(\phi)$ 
[$\gamma^\infty(\phi) = 0$ for half-band filling]. 
An analogous scaling behavior has been found in 
previous numerical studies of $W(\gamma_{12})$ of non-dimerized 
Hubbard models, where $\gamma_{12}$ refers to the NN density-matrix 
element.\cite{prbxcfun} In this case one observes that $W(\gamma_{12})$
depends weakly on system size $N_a$, band-filling $n = N_e / N_a$, 
and lattice structure, if $W$ is measured in units of the Hartree-Fock 
energy $E_{\rm HF}$ and if $\gamma_{12}$ is scaled within the relevant 
domain of representability $[\gamma_{12}^\infty, \gamma_{12}^0]$.
In the present context, Fig.~\ref{fig:Wgren} implies that the change 
in $W$ associated to a given change in the degree of delocalization 
$\gamma / \gamma^0(\phi)$ can be regarded as nearly independent of 
$\phi$ and system size. A good general approximation to 
$W(\gamma, \phi)$ can then be obtained by applying such a scaling 
to the functional dependence extracted from a simple reference system. An 
appropriate choice is provided by the fully-dimerized chain
corresponding to $\phi=0$, which can be worked out analytically. 
In this case the system consists of a collection of dimers
and the exact interaction energy reads
\begin{equation}
\label{eq:Wphi0}
W(\gamma, \phi \! = \! 0) = {U N_a\over 4}
\left( 1  - \sqrt{ 1 -  \gamma^2}\right) \; .
\end{equation} 
Scaling the functional dependence of the dimer interaction energy 
to the $\phi$-dependent domain of representability one obtains
\begin{equation}
\label{eq:Wdim}
W_0(\gamma,\phi) = {U N_a\over 4}
\left( 1  - \sqrt{ 1 -  \left[{\gamma \over \gamma^0(\phi)}\right]^2 } 
\right) \; ,
\end{equation} 
which we propose as approximation to $W$ for dimerized systems. Notice 
that $W_0(\gamma,\phi)$ preserves the previous general properties 
(i)--(iv) and that it is of course exact for $\phi =0$
[$\gamma^0(\phi\!=\!0) = 1$]. In practice, the system specific 
function $\gamma^0(\phi)$ can be easily obtained by integration of the 
single-particle spectrum.

It is important to remark that the density matrices $\gamma_{ij}$ 
involved in the approximate functional $W_0$ are pure-state 
$N$-representable. Eq.~(\ref{eq:Wdim}) applies to the $\gamma_{ij}$ 
obtained by scaling the off-diagonal elements of the density matrices 
$\gamma_{ij}^0$ that derive from uncorrelated states 
$|\Psi_0\rangle$ having $N_e$ electrons on $N_a$ sites, and 
a uniform density distribution 
$\langle \Psi_0| \sum_{\sigma} \hat n_{i \sigma} | \Psi_0\rangle 
= N_e/N_a = 1$. In other terms, $\gamma_{ij}$ has the form
$\gamma_{ij} = \lambda \gamma_{ij}^0$ with 
$0\le \lambda\le 1$ for all $i\not= j$, and 
$\gamma_{ii} = \gamma_{ii}^0 = 1$ for all $i$.
In order to show the pure-state representability of $\gamma_{ij}$,
we consider two normalized $N$-particles states
$|\Psi_a\rangle$ and $|\Psi_b\rangle$ satisfying 
$\langle\Psi_a| \sum_{\sigma} \hat c_{i \sigma}^{\dagger} \hat c_{j \sigma}
|\Psi_b\rangle = 0$ for all $ij$. This condition is fulfilled, for example, 
by states $a$ and $b$ having different defined total spins $S$ or $S_z$, 
or by superpositions of pure-$S$ or pure-$S_z$ states 
sharing no common eigenvalues. The density-matrix represented by 
$|\Psi\rangle = \alpha |\Psi_a\rangle + \beta |\Psi_b\rangle$ with
$\alpha^2 + \beta^2 = 1$ is then given by 
$\gamma_{ij} = \langle\Psi| 
\sum_{\sigma} \hat c_{i \sigma}^{\dagger} \hat c_{j \sigma}|\Psi\rangle=
\alpha^2 \gamma_{ij}^{a} + \beta^2 \gamma_{ij}^{b}$, 
where $\gamma_{ij}^{a}$ and $\gamma_{ij}^{b}$ are the density matrices
corresponding to $|\Psi_a \rangle$ and $|\Psi_b \rangle$. 
Therefore, all the density matrices in the segment defined by 
$\gamma_{ij}^{a}$ and $\gamma_{ij}^{b}$ are pure-state $N$-representable.
The representability of a scaled uncorrelated 
$\gamma_{ij}^0$ at half-band filling follows from the previous lemma
by taking $|\Psi_a \rangle = |\Psi_0\rangle$, which has $S = 0$ or $1/2$, 
and $|\Psi_b \rangle$ equal to the fully localized state with 
one electron per site and maximal $S= N_e/2$,
for which $\gamma_{ij}^{b}=0$ for all $i \not = j$,
and $\gamma_{ii}^{b}=1$ for all $i$. 
Consequently, the $\gamma_{ij}$ in the domain of definition of 
$W_0$ and the ground-state density matrices $\gamma_{ij}^{gs}$ 
derived from it are all pure-state $N$-representable. 

\begin{figure}
\includegraphics[scale=0.75]{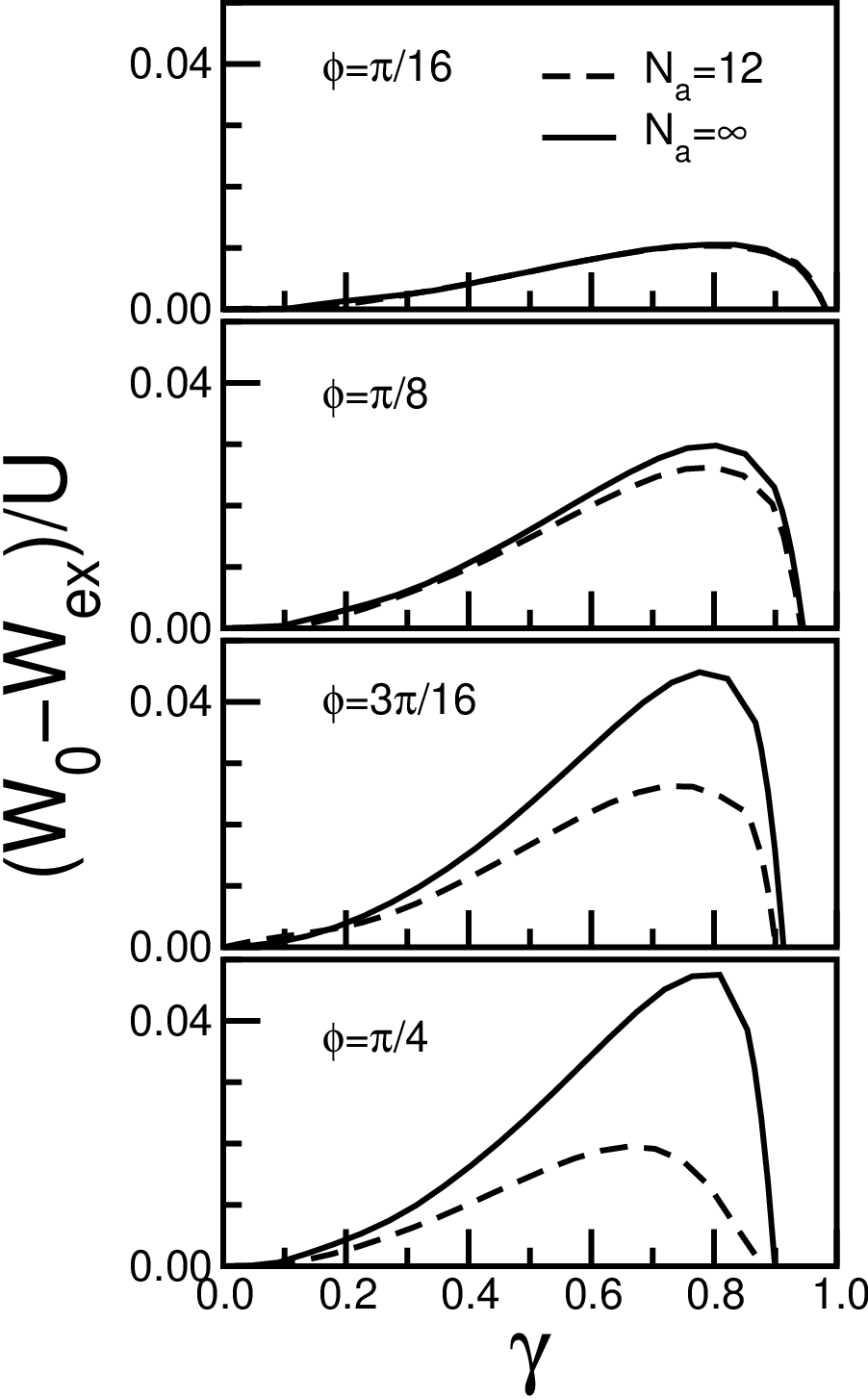}
\caption{\label{fig:Wdif}
Comparison between the exact interaction-energy functional $W_{ex}$ 
of the Hubbard model and the approximation $W_0$ given by 
Eq.~(\protect\ref{eq:Wdim}). Results are given for the $N_a = 12$ 
site ring (dashed) and the infinite chain (solid) as a function of 
$\gamma = \sqrt{\gamma_{12}^2 + \gamma_{23}^2}$ 
for different $\phi= \arctan(\gamma_{12}/\gamma_{23})$.
        }
\end{figure}

Fig.~\ref{fig:Wdif} compares Eq.~(\ref{eq:Wdim}) with the exact 
$W(\gamma,\phi)$ for a $12$-site Hubbard ring and for the infinite 
chain. One observes 
that the proposed approximation follows rather closely the exact 
results for all $\gamma$ and $\phi$. The largest discrepancies 
are found for vanishing or moderate dimerization 
(e.g., $\phi = 3\pi/16$ or $\phi = \pi/4$) and relatively large $\gamma$
($\gamma \simeq 0.8$). In all cases the quantitative differences remain
small ($|W_0 - W| / U \le 0.047$ for $\phi = \pi/4$ and 
$|W_0 - W| / U \le 0.045$ for $\phi = 3\pi/8$) which is quite 
remarkable taking 
into account the simplicity of the approximation. In the following, 
Eq.~(\ref{eq:Wdim}) is applied in the framework of LDFT to 
determine several properties of the dimerized 1D Hubbard model.
Comparison is made with exact results whenever possible in order 
to assess the performance of the method.

\section{\label{sec:dimhub}Dimerized Hubbard chains} 

\begin{figure}
\includegraphics[scale=0.73]{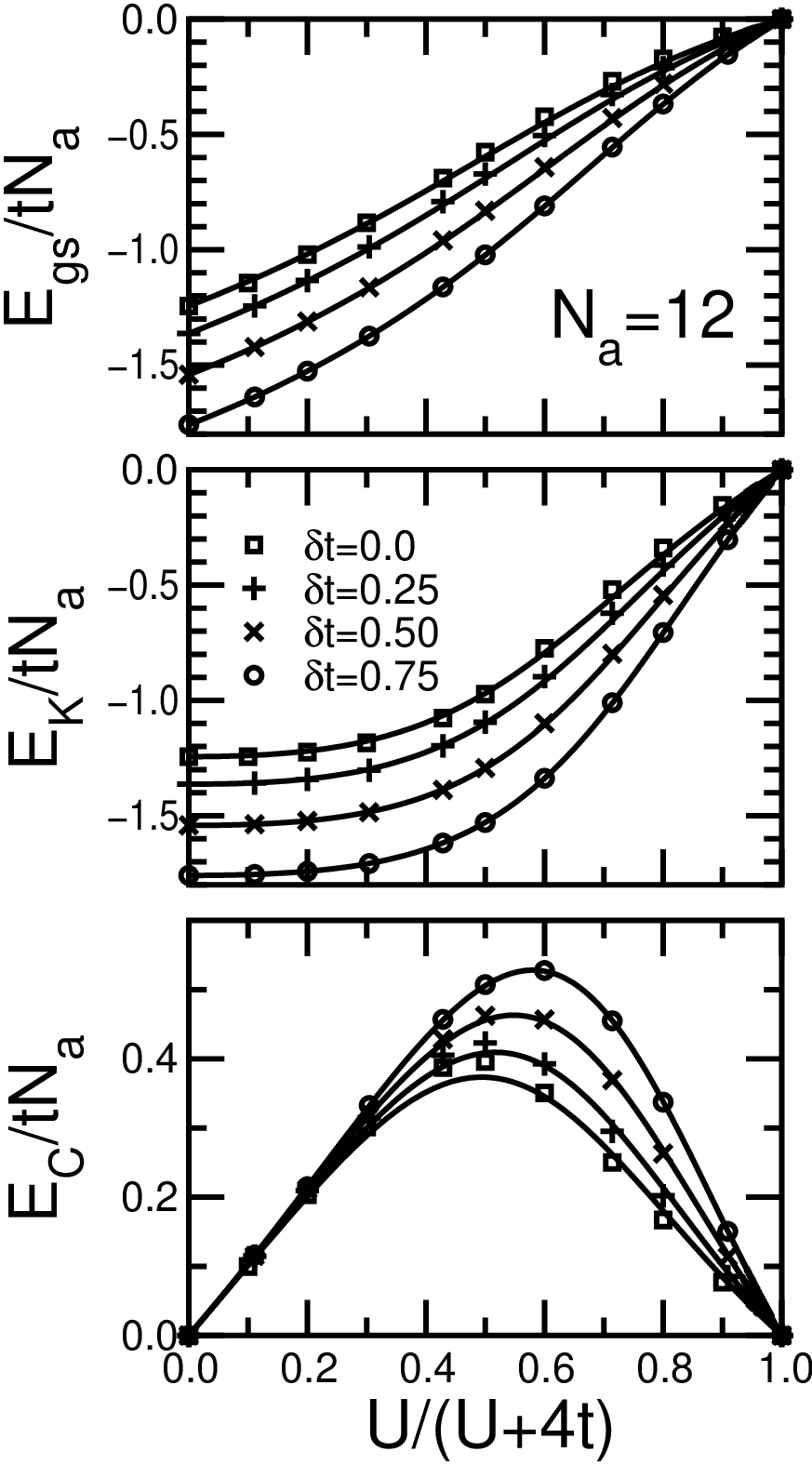}
\caption{\label{fig:E12}
Ground-state energy $E_{gs}= E_{K} + E_{C}$, kinetic energy $E_{K}$, 
and Coulomb energy $E_{C}$ of dimerized Hubbard rings with hopping 
integrals $t_{ij} = t (1 \pm \delta t)$, Coulomb interaction $U$, 
$N_a=12$ sites, and $N_e = N_a$ electrons. The symbols are obtained 
from exact Lanczos diagonalizations\protect\cite{lanczos} 
and the solid curves correspond to the present lattice 
density-functional theory.
        }
\end{figure}

\begin{figure}
\includegraphics[scale=0.73]{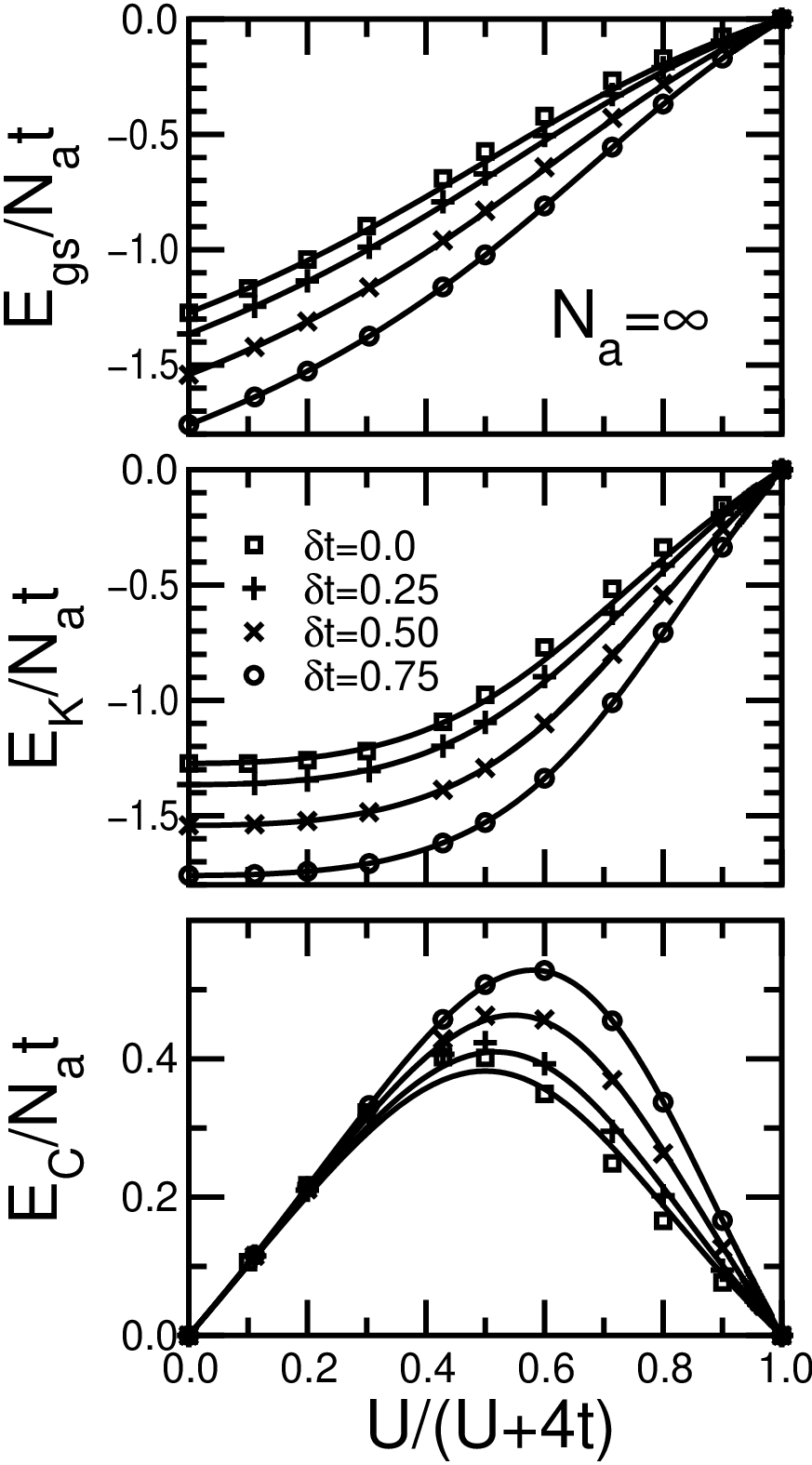}
\caption{\label{fig:Einf}
Ground-state energy $E_{gs}= E_{K} + E_{C}$, kinetic energy $E_{K}$, 
and Coulomb energy $E_{C}$ of dimerized infinite Hubbard chains with 
hopping integrals $t_{ij} = t(1 \pm \delta t)$. The symbols are 
obtained using the density-matrix renormalization group 
method\protect\cite{white} and the solid curves correspond to 
the present lattice density-functional theory 
(see Fig.~\protect\ref{fig:E12}).
        }
\end{figure}

In Figs.~\ref{fig:E12} and \ref{fig:Einf} the ground-state energy 
$E_{gs}$, kinetic energy $E_{K}$, and Coulomb 
energy $E_{C}$ of the 1D Hubbard model are given as a function 
of the Coulomb repulsion strength $U/t$ for different hopping 
alternations $\delta t$. 
Accurate numerical results are also shown which were obtained 
using the Lanczos diagonalization method\cite{lanczos} for $N_a = 12$ 
or the DMRG method\cite{white} for the infinite chain. In the case 
of the non-dimerized infinite chain, Lieb and Wu's exact 
solution\cite{lieb-wu} is taken as reference. The results for $N_a = 12$ 
and $N_a= \infty$ are qualitative very similar. 
$E_{gs}$ increases monotonically with $U/t$ since
$\partial E_{gs}/\partial U = 
\langle \hat n_{i\uparrow} \hat n_{i\downarrow}\rangle > 0$, 
vanishing in the limit of $U/t = \infty$. For $U/t < 4$ this 
is essentially a consequence of the increase of $E_C \propto U$, as
$E_K$ and $\gamma_{ij}$ remain very much like in the uncorrelated 
$U=0$ state. In contrast, for $U/t > 4$ the electrons become 
increasingly localized, and the increase of $E_{gs}$ results form
the increase of $E_K$ which approaches zero as $|\gamma_{ij}|$ decreases.
At the same time $E_C$ tends to zero as charge fluctuations are suppressed
(see Figs.~~\ref{fig:E12} and \ref{fig:Einf}).

Comparison between LDFT and the exact results shows a very good 
agreement. This concerns not only $E_{gs}$ but also the separate 
kinetic and Coulomb contributions indicating that electron localization
and intra-atomic correlations are correctly described for all $U/t$.
Moreover, this also shows that the results obtained for the 
ground-state energy are not the consequence of strong compensations 
of errors. Concerning the accuracy of $E_K$ and $E_C$ one generally
observes that a somewhat higher precision is achieved for $E_K$, 
which functional dependence is known exactly, as compared to $E_C$, 
which derives from an approximation to $W$ [Eq.~(\ref{eq:Wdim})]. 
For $\delta t/t \ge 0.1$ the LDFT calculations are nearly 
indistinguishable from the exact ones (e.g., 
$|E_{gs} - E_{gs}^{ex}|/t \le 0.03$ for $\delta t / t = 0.1$). 
Even the largest quantitative discrepancies, found for the non-dimerized 
chain at intermediate $U/t$, are pretty small. For instance,  
for $\delta t = 0$ and $U/t = 4$ we obtain 
$|E_{gs} - E_{gs}^{ex}|/t = 0.020$ for the 12-site ring and 
$|E_{gs} - E_{gs}^{ex}|/t =0.044$ for the infinite chain. 
Comparing Figs.~\ref{fig:E12} and \ref{fig:Einf} one observes 
that the performance of the method is sometimes higher for the 
12-site ring than for the infinite chain. For example, 
Fig.~\ref{fig:Einf} shows that $E_K$ ($E_C$) is slightly 
overestimated (underestimated) for $\delta t = 0$ and 
$U/(U+4t) = 0.7$--$0.8$, whereas for $N_a = 12$ a much better
agreement with the exact result is found (see Fig.~\ref{fig:E12}). 
In any case it is important to recall that no artificial 
symmetry breaking is required to describe correlation-induced 
localization correctly, as it often occurs in other approaches 
(e.g., mean-field spin-density-wave state). Moreover, the present 
calculations remain simple and numerically not demanding, 
since the minimization of $E[\gamma_{ij}]$ is performed 
using analytical expressions for $E_K$ and $W$ 
[see Eqs.~(\ref{eq:ek}) and (\ref{eq:Wdim})]. One concludes that 
LDFT, combined with Eq.~(\ref{eq:Wdim}) as approximation to the 
interaction energy functional, provides an efficient and correct 
description of the ground-state properties of the 1D Hubbard model 
in the complete range of interaction strength and dimerization. 

\begin{figure}
\includegraphics[scale=0.75]{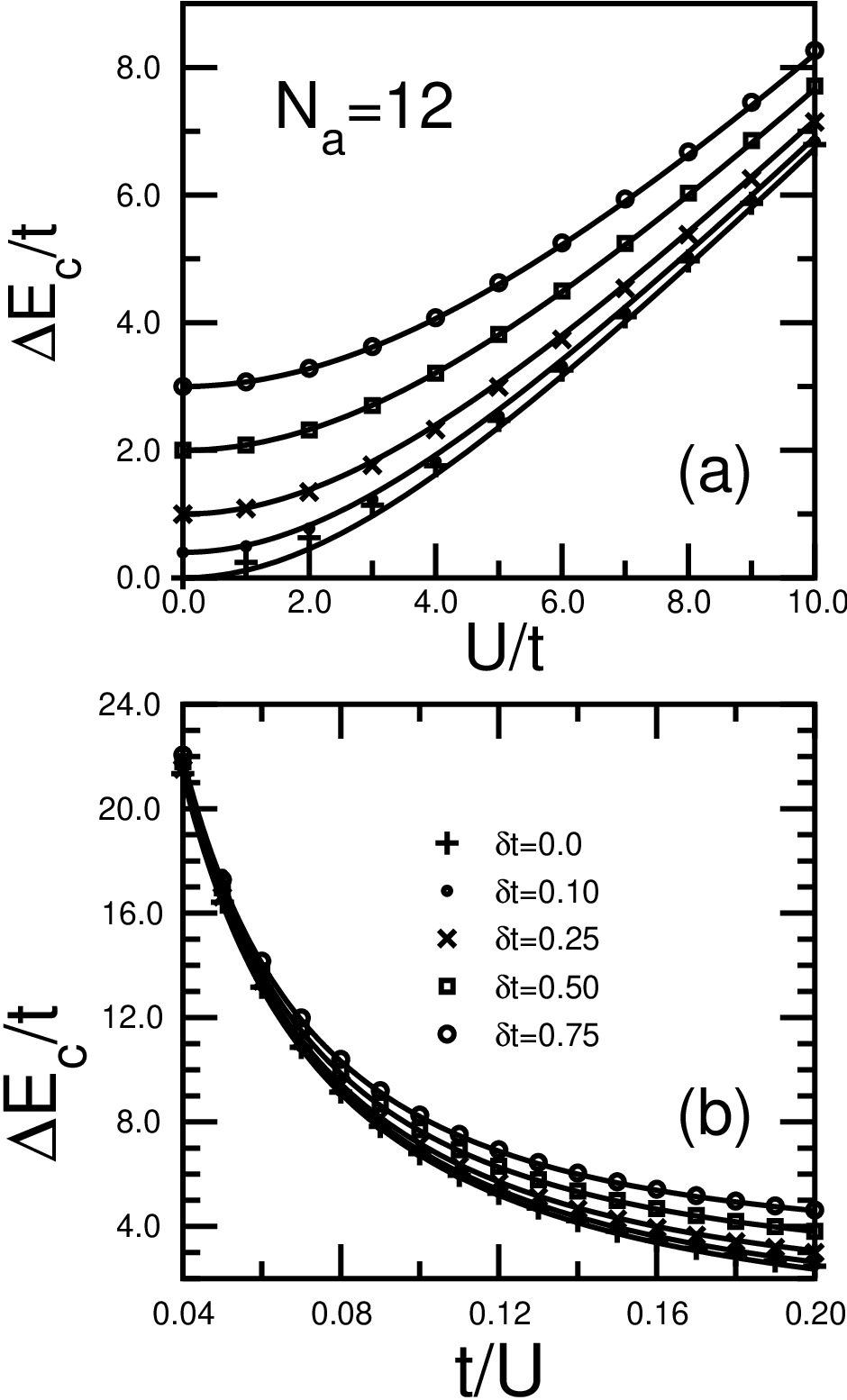}
\caption{\label{fig:hubgap12}
Charge-excitation gap $\Delta E_c$ of dimerized Hubbard rings 
with $N_a = 12$ sites, hopping integrals $t_{ij} = t(1 \pm \delta t)$, 
and band filling $n = N_e/N_a = 1$. The symbols refer to exact 
numerical diagonalizations\protect\cite{lanczos} 
and the curves to the present lattice density-functional approach.
         }
\end{figure}

\begin{figure}
\includegraphics[scale=0.75]{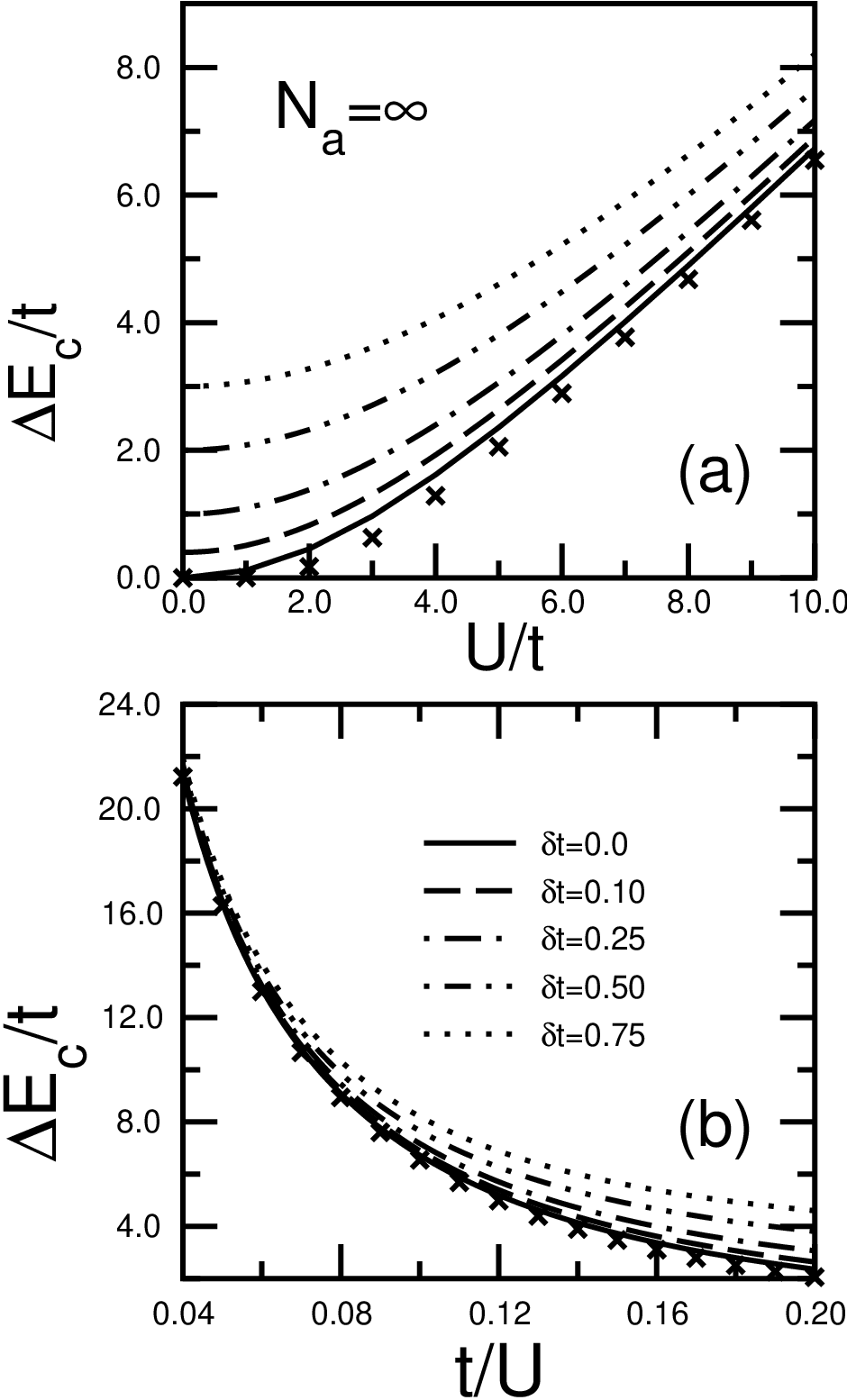}
\caption{\label{fig:hubgapinf}
Charge-excitation gap $\Delta E_c$ of dimerized 1D Hubbard chains 
with hopping integrals $t_{ij} = t(1 \pm \delta t)$ and band filling 
$n = N_e/N_a = 1$. The crosses refer to exact Bethe-Anstaz results for 
$\delta t = 0$ (see also Fig.~\protect\ref{fig:hubgap12}). 
         }
\end{figure}

The charge-excitation or band gap
\begin{equation}
\Delta E_c =  E_{gs}(N_{e}+1) + E_{gs}(N_{e}-1) - 2E_{gs}(N_{e})
\end{equation}
is a property of considerable interest in strongly correlated systems
which can can be related to the discontinuity in the derivative of 
the kinetic and correlation energies per site with respect to the 
electron density $n$. The determination of $\Delta E_c$ constitutes 
a much more serious challenge than the calculation of ground-state 
properties like $E_{gs}$, $E_K$, and $E_C$ particularly in the 
framework of a density-functional formalism. Results for 
$\Delta E_c$ of the 1D Hubbard model are given in
Figs.~\ref{fig:hubgap12} and \ref{fig:hubgapinf} as a function 
of the Coulomb repulsion strength $U/t$ for different values of 
the hopping alternation $\delta t$ ($n=1$). $\Delta E_c$ vanishes 
for $\delta t = 0$ and $U/t = 0$, and increases with increasing 
$U/t$ or $\delta t$. Comparison between LDFT and 
Lanczos exact diagonalizations ($N_a = 12$) or the
Bethe-Ansatz solution\cite{lieb-wu} ($N_a = \infty$ and 
$\delta t = 0$) shows fairly small quantitative discrepancies. 
In the most difficult non-dimerized case we find 
$|\Delta E_c - \Delta E_c^{ex}| < 0.18t$ for $N_a = 12$, and 
$|\Delta E_c - \Delta E_c^{ex}| < 0.34t$ for $N_a = \infty$. 
For small $U/t$ and $\delta t = 0$, $\Delta E_c$ is somewhat 
underestimated for $N_a = 12$ and overestimated for $N_a = \infty$.
The latter is mainly due to the fact that 
Eq.~(\ref{eq:Wdim}) fails to reproduce the exponential decrease 
of $\Delta E_c$ for $U/t \to 0$ ($N_a = \infty$ and 
$\delta t = 0$).\cite{lieb-wu} 
As in previous properties the accuracy improves with increasing 
$\delta t$. Fig.~\ref{fig:hubgap12} shows that the LDFT results 
for non-vanishing dimerization and $N_a = 12$ are very close 
to the exact ones ($|\Delta E_c - \Delta E_c^{ex}| / t < 0.011$
already for $\delta t/t = 0.1$). Therefore, one expects that the 
predictions for $N_a = \infty$ and $\delta t > 0$ should be reliable. 
Finally, one may note that in the limit of large $U/t$, the hopping 
alternation $\delta t$ has little effect on the charge gap. 
As the electrons tend to localize for
$U/t \to \infty$, $\Delta E_c \to U + E_b$ where $E_b = -4t$ is the 
energy of the bottom of the single-particle band. The present 
lattice-density-functional scheme describes correctly the crossover 
from a band insulator to a Mott insulator which occurs in dimerized 
chains as $U/t$ is varied from the weak-interaction 
to the strong-interaction regime.

\section{\label{sec:concl}Summary and Outlook} 

A novel density-functional approach to lattice-fermion models has been 
applied to the dimerized 1D Hubbard Hamiltonian. In this framework
the basic variable is the single-particle density matrix $\gamma_{ij}$ 
and the key unknown is the interaction-energy functional $W[\gamma_{ij}]$.
In the present paper we have first investigated the functional dependence 
of $W$ on the density-matrix elements $\gamma_{12}$ and $\gamma_{23}$
between nearest neighbors in dimerized chains
($\gamma_{i,i+1}=\gamma_{12}$ for $i$ odd and 
$\gamma_{i,i+1}=\gamma_{23}$ for $i$ even). Rigorous results for
$W(\gamma_{12}, \gamma_{23})$ were derived from 
finite-ring Lanczos diagonalizations and from DMRG calculations 
for the infinite chain. An analysis of these exact results shows
that $W$ can be appropriately scaled as a 
function $\gamma / \gamma^0(\phi)$, where 
$\gamma = \sqrt{\gamma_{12}^2 + \gamma_{23}^2 }$ and
$\gamma^0(\phi)$ is the largest representable $\gamma$ for a given
$\phi = \arctan(\gamma_{12}/ \gamma_{23})$.
A simple general approximation to $W$ was then proposed which takes 
advantage of this scaling behavior and which provides with a unified
description of correlations from weak to strong coupling regimes. 
Finally, using this approximation, several ground-state properties
and the charge-excitation gap of dimerized chains have been 
determined successfully as a function of Coulomb repulsion strength 
$U/t$ and hopping alternation $\delta t$. 

The accuracy of the
results encourages more or less straightforward applications 
of the present approach to related problems such as multi-leg 
ladders or the two-dimensional square lattice with first 
and second NN hoppings ($t$-$t'$ Hubbard model). Moreover, 
the possibility of generalizing the present scaling approximation 
to an arbitrary number of independent variables $\gamma_{ij}$
deserves to be investigated in detail, since it would open the way 
to applications in very low symmetry situations including metal 
clusters and disordered systems.

\begin{acknowledgments}

Helpful discussions with Dr.\ Ph.\ Maurel are gratefully acknowledged.
This work has been supported by CONACyT (Mexico) through the project
W-8001 (Millennium initiative).
Computer resources were provided by IDRIS (CNRS, France).

\end{acknowledgments}

\vfill\break

%
%

%


\begin{thebibliography}{}
\bibitem {foot_ens}
An extension of the definition domain of $E[\gamma_{ij}]$ to 
ensemble-representable density matrices $\Gamma_{ij}$ is straightforward 
following the work by Valone (see Ref.~\protect\cite{val}). 
Ensemble density-matrices are written as $\Gamma_{ij} = \sum_n w_n 
\langle\Psi_n| \sum_\sigma \hat c_{i \sigma }^{\dagger} \hat c_{j \sigma }
|\Psi_n \rangle$ with $w_n \ge 0$ and $\sum_n w_n = 1$. 
In practice, they are much easier to characterize than 
pure-state density matrices.
\end{thebibliography}
\end{document}